\documentclass[12pt]{article}
\usepackage{rotate}
\usepackage{epsfig}
\usepackage{psfrag}
\usepackage{a4}
\usepackage{graphicx}
\begin{document}

\title{A Crash Course on Aging}
\author{Giulio Biroli}
\date{\ }
\maketitle
\vspace{-0.7in}
\begin{center}
{\small \it
Service de Physique Th{\'e}orique, Orme des Merisiers, CEA Saclay,
91191 Gif sur Yvette Cedex, France.}
\end{center}
\vspace{0.35in}

\begin{abstract}
In these lecture notes I describe some of the main theoretical ideas
emerged to explain the aging dynamics.
This is meant to be a very short introduction to aging dynamics and no
previous knowledge is assumed. I will go through simple examples 
that allow one to grasp the main results and predictions. 
\end{abstract}
\newpage
\section{Contents}
As the title makes it clear, the aim of these lectures is not to present 
an exhaustive and complete description of the state of the art in 
aging dynamics (even less an exhaustive reference list).
But rather, its purpose is to present in a nutshell the main 
theoretical ideas emerged to explain the aging dynamics.   
For a more complete and deep presentation there are already very good 
reviews and lectures in the literature\cite{ROE,ReviewSaclay}. 
These references are also useful to have an introduction to 
the experimental results on aging (in particular the ones
on memory and rejuvenation effects) which will not be
discussed in the following.
\vskip 1cm

The contents of these lectures are:{\bf
\begin{itemize}
\item Introduction to aging and off-equilibrium thermal relaxation
\item A simple example: domain growth in the Ising model
\item Mean-Field Theory of Aging
\item Activated dynamic and scaling
\item The Trap model
\item  Outstanding issues
\end{itemize}
}

\newpage
\section{Introduction}\label{introduction  to aging and off-equilibrium thermal relaxation}
\begin{figure}
\centerline{\epsfxsize=10cm
\epsffile{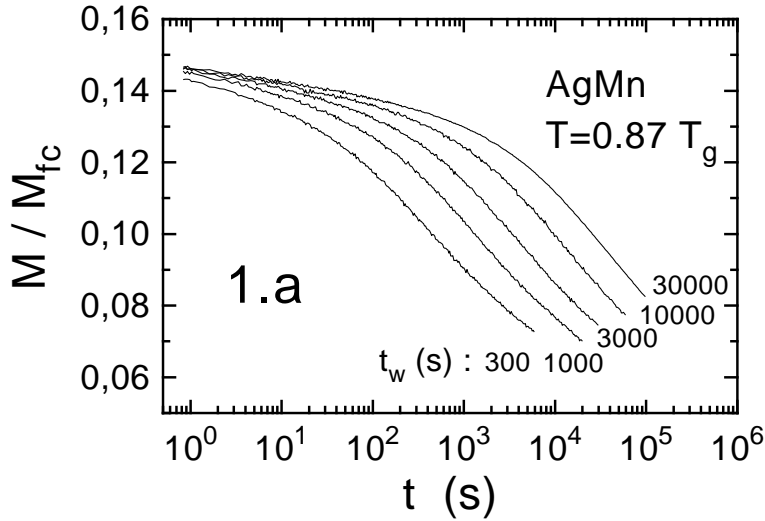}
}
\vspace{.2cm}
\caption{\label{fig1}
Thermo-remanent magnetization $M$, normalized by the field-cooled value
$M_{fc}$,
{\it vs.}~$t (s)$ ($\log_{10}$ scale) for a spin-glass 
($Ag:Mn_{2.6\%}$ sample), at $T=9 K=0.87 T_c$. The sample has been 
cooled in a $0.1\ Oe$ field from above
$T_g=10.4K$ to 9K; after waiting $t_w$, the field has been cut at $t=0$, and
the decaying magnetization recorded (from \cite{ReviewSaclay}).}
\end{figure}

Off-equilibrium dynamics is a very wide and interesting subject. Many
works have been devoted to it recently, new interesting theoretical
concepts as well as fascinating experimental results have been discovered.
Yet, it is certainly fair to say that the overall comprehension is
still not complete and certainly many new discoveries (as well as much
work) is ahead. There are roughly two different types of off-equilibrium
dynamics: driven off-equilibrium dynamics and thermal off-equilibrium
relaxation. In the first case the system is driven out of equilibrium by
an external stationary force (it may be a shear for a liquid, a
voltage for an electron system, etc\dots ) and kept in a stationary off-equilibrium
state. The subject of these lectures is the latter type of
off-equilibrium dynamics: thermal off-equilibrium
relaxation. In this case the system, at equilibrium at time
$t_{0}$, is brought out of equilibrium by changing a control parameter $C$ (for
example the magnetic field for a magnetic system, or the temperature
for a glass-forming liquid, etc\dots ). Subsequently the system, which is in
general in contact with a thermal bath, starts to evolve towards the new equilibrium state corresponding
to the new value of $C$. However, this may take a long time which
actually can become infinite in the thermodynamic limit or just for
practical purposes. \\
From a
theoretical point of view this is a clear
example in which ``more is different''. Indeed a system with few
degrees of freedom coupled to a thermal bath is going always to
equilibrate in a finite time\footnote{We neglect the cases in which the configuration
spaces of the system with few degrees of freedom is broken in
disconnected pieces.} (which in the worst case scenario and for
the majority of systems scales with the number $N$ of the degrees of freedom as $e^{NK}$ where
K is a positive constant). However, when the number degrees of freedom is very large
this equilibration time may become infinite for all practical purpose.
As in standard equilibrium statistical mechanics the way to deal with
this situation is study the system taking the thermodynamic limit from the beginning thus focusing on the evolution of the system on 
timescales which may be extremely large but do not diverge with the
system size. \\
Let us give just an example before introducing the aging
dynamics. Consider a three dimensional 
Ising ferromagnet at a temperature $T$ larger
than the critical temperature $T_{c}$. If one quenches the system from
$T$ to $T'>T_{c}$ the ferromagnet will relax (i.e. its energy, its
local correlation functions, etc\dots , will relax) on a finite timescale
to the new equilibrium state at temperature $T'$. However, 
it has been proven that for $T'<T_{C}$ the largest relaxation time scales
as $\exp K N^{2/3}$, hence, in the thermodynamic limit, 
the system remains out of equilibrium forever
(in the next section we are going to understand
in detail the properties of the off-equilibrium dynamics in this case).
Note that this never-ending off-equilibrium regime 
is quite general whenever the quench crosses a phase
transition. Let us consider another example: the measurement of
thermo-remanent magnetization in a spin glass. In this case the system
is cooled in a small field from above $T_{c}$ to a temperature below, 
it then waits in the field for a time $t_{w}$ after which the field is
switched off. The subsequent evolution of the magnetization is very
peculiar and an example is reported in Fig. \ref{fig1}.
 After a very rapid fall-off
of the magnetization there is a very slow decreasing. A very important
feature of these evolutions is that the longer is $t_{w}$ the slower is
the relaxation. This behavior is called aging
because the system evolution depends on its age ($t_{w}$). This means that
the typical relaxation timescale for the system is not fixed a priori
but is evolving and is fixed by the age of the system itself: older systems
relax more slowly.\\
In order to understand better this point let us again focus on a system
with a {\it finite number}  of degrees of freedom [which therefore has 
a {\it finite} relaxation time]. Generically one expects a
similar aging behavior for not too large $t_{w}$. However when the
typical relaxation time scale at time $t_{w}$ becomes of the order 
of the equilibration time 
all the aging effects are interrupted 
and the system is characterized by an equilibrium dynamics.
This in particularly means that one time quantities like the
magnetization, the energy, etc. equal their equilibrium values
and, hence, they are time independent; whereas for example 
the two-time correlation and response functions are invariant under 
translation of time, i.e. they depend just on the time difference.
Finally typical equilibrium relations like fluctuation dissipation
relations between correlation and response are verified.\\
Thus, it is clear that aging and the never-ending relaxation 
discussed before is a collective phenomenon
with (at least one) growing correlation length
\footnote{Why? Well, it is the usual sloppy argument. If all the correlation
lengths are bounded during the relaxation then one can roughly divide the system
in independent sub-systems for which the relaxation time is bounded by
a certain timescale $t_{B}$. Thus the aging has to be interrupted and
the system has to relax on a timescales not larger than $t_{B}$ which
contradicts the hypothesis of never reaching equilibrium 
on any finite timescales.
In real systems this argument can be used to see that one should have
a growing (although not diverging) correlation length even for systems 
with interrupted aging. Finally, a word of caution, this argument 
might have problems for system with long range interactions,
mean-field models and quench to zero temperature.}.
Indeed one of the theoretical approaches to understand and explain
aging dynamics is based on determining what is the growing correlation 
length, how it grows with time and assuming scaling with respect to
this length. Another approach is based on the analytical solution 
of infinite dimensional (or also infinite number of components)
models. In this case an interpretation in terms of a growing
lengthscale is at a first sight indirect. 
Instead the standard physical interpretation that has been developed
during the last ten years is in terms of evolution within the energy
landscape.\\
In the following we shall analyze the aging dynamics in the Ising
model. This is a good example to start with because it is rather simple
and allows one to introduce both approaches which 
are correct and complementary (in this case at least!). 
In the following sections I will
present more in detail both approaches focusing in particular
on their application to disordered systems. Finally I will also
briefly present the trap model, that although certainly not a microscopic
model, is used often in phenomenological studies and appears as a coarse
grained description for some microscopic models.

\section{A simple example: domain growth in the Ising model}\label{Ising}
\begin{figure}
\centerline{\epsfxsize=10cm
\epsffile{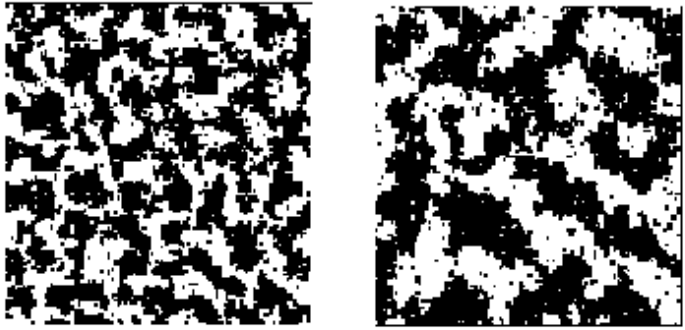}
}
\vspace{.2cm}
\caption{\label{fig2}
Two snapshots of a 2d cut of a 3d ferromagnetic Ising model evolving 
with a Glauber dynamics after a quench at time $t=0$ from high temperature to
$T<T_{c}$. Black color is used for minus spins and white for plus
spins. On the left the configuration reached after $t_{w}=10^{3}$
MonteCarlo steps. On the right the configuration reached after $t_{w}=10^{5}$
MonteCarlo steps. Statistically the two configurations looks the same 
after a length rescaling. 
}
\end{figure}

In the following I shall focus on the aging dynamics of the three
dimensional Ising model
due to a quench from the high disordered temperature phase to a temperature
at which, in equilibrium, the system is ferromagnetically ordered. 
This example is particularly enlightening because it can be analyzed
in a simple approximated way 
(of course much more sophisticated analyzes have been performed). 
Furthermore all the different theoretical ways to tackle the
aging dynamics can be introduced and their physical content can be
easily understood.\\
The first thing to do in order to understand what is going on is to
look at the results of a numerical simulations. In Fig. \ref{fig2} 
there are two snapshots of a 2d cut of a 3d ferromagnetic Ising model
after a quench from the high to the low temperature phase 
(black color is used for minus spins and white for plus
spins). This figure clearly shows that the systems is 
trying to separate minus and plus regions in order to gain bulk
energy and decrease interface energy: the system is separated in
domains that grow in time.\\
In order to analyze this off-equilibrium dynamic regime
it is convenient to set up a continuum description in terms of the
coarse-grained magnetization density $\phi ({\bf x},t)$ which obeys a
Langevin equation:
\begin{equation}\label{Langevin}
\partial_{t} \phi ({\bf x},t)=-\frac{\partial F}{\partial \phi ({\bf
x},t)}+\eta ({\bf x},t)\quad , 
\end{equation}
where $\eta ({\bf x},t)$ is a white noise with a variance determined
by the temperature $T$
\[
<\eta ({\bf x},t)\eta ({\bf
x}',t')>=2T\delta ({\bf x}-{\bf x}')\delta (t-t')
\]
and $F$ is the standard Ginzburg-Landau free energy
\begin{equation}\label{F}
F=\int d{\bf x} \left[ \frac{1}{2} (\nabla \phi
)^{2}-\frac{m^{2}}{2}\phi^{2}+\frac{g}{4}\phi^{4}\right]\quad .
\end{equation}
The other important information that one has to provide is the initial
condition that we are going to take of the form:
\begin{equation}\label{ic}
<\phi  ({\bf x},0)\phi  ({\bf
x}',0)>=\Delta \delta ({\bf x}-{\bf x}')\quad .
\end{equation}
Since we are focusing on a system that is in equilibrium at a very high
temperature at $t=0$, the correlation length is extremely small and,
at our coarse-grained level, the magnetization density correlations
are just given by (\ref{ic}). At $t=0^{+}$ the bath temperature 
is instantaneously switched to $T<T_{c}$, where $T_{c}$ is the critical
temperature of the Ising model. Hence, the system starts to evolve 
trying to thermalize without never succeeding on any finite time. \\
In this case the temperature $T$ does not play a very
important role, as long as it is smaller than $T_{c}$. Therefore we
shall put it equal to zero and we will discuss later what happens for
non zero temperature. \\
As a consequence we have to solve the differential equations:
\begin{equation}\label{t0}
\partial_{t} \phi ({\bf x},t)= (\triangle +m^{2})\phi ({\bf x},t)-g\phi ({\bf x },t)^{3}
\end{equation}
with the initial conditions (\ref{ic}). These eqs. correspond to a pure
gradient descent in the energy landscape defined by $F$. 
There are two terms in $F$: the bulk-energy term
$-\frac{m^{2}}{2}\phi^{2}+\frac{g}{4}\phi^{4}$ that is minimized by $\phi=\pm
\sqrt{m^{2}/g}$ and the $(\nabla \phi
)^{2}$ which is minimized by flat configurations.
As a consequence the system during its gradient descent will try to
decrease $F$ and this will lead to fatter and flatter regions
 with a $\phi=\pm
\sqrt{m^{2}/g}$ as seen in Fig. \ref{fig2}.\\
Equations (\ref{t0}) are somewhat simpler than
the original stochastic ones but they are still too complicated to be
analyzed exactly. Therefore in the following we are going to make use
of an approximation {\`a} la Hartree that can be justified as a $1/N$
expansion. It consists simply in replacing the nonlinear term 
$g\phi ({\bf x},t)^{3}$ by $g<\phi ({\bf x},t)^{2}>\phi ({\bf x},t)$
where $<\cdot>$ means the average over the initial condition
\footnote{If one generalizes the field $\phi ({\bf x},t)$ to an
n-components field $\phi^{\alpha }({\bf x},t)$ and F to 
$\int d{\bf x} \left[ \frac{1}{2} (\nabla \vec{\phi}
)^{2}-\frac{m^{2}}{2}\vec{\phi}^{2}+\frac{g}{4N}
(\vec{\phi}^{2})^{2}\right]$ then the equation (\ref{t0}) becomes
exact in the infinite $N$ limit. The reason is that in
eq. (\ref{t0}) the term $g\phi^{3}$ becomes 
$g\phi^{\alpha } \vec{\phi}^{2}/N$ and in the infinite $N$ limit the
term $\vec{\phi}^{2}/N$ does not fluctuate and is equal to its
average.}. Now the equations read simply:
\begin{equation}\label{simple}
\partial_{t} \phi ({\bf x},t)= [\triangle +a (t)]\phi ({\bf
x},t)
\end{equation}
where $a (t)=m^{2}-g<\phi ({\bf x},t)^{2}>$ has to be determined
self-consistently (note that since the initial average is translation
invariant $<\phi ({\bf x},t)^{2}>$ does not depend on ${\bf x}$).
Eq. (\ref{simple}) can be immediately integrated in Fourier space:
\[
\phi ({\bf k},t)=\phi ({\bf k},0)\exp \left(-k^{2}t+\int_{0}^{t}a (t')dt' \right)
\]

and the self-consistent equation on $a (t)$ reads:
\[
a (t)=m^{2}-g\Delta \int \frac{d{\bf k}}{(2\pi)^{3}}\exp \left(-2 k^{2}t+2 \int_{0}^{t}a (t')dt' \right)
\]
where eq. (\ref{ic}) has been used to eliminate the initial condition
\footnote{Some readers can find that this expression is a bit singular
at $t=0$. In this case one has to remember that there is an underlying
lattice and there is a cut-off in the integration over $k$ corresponding
to the inverse of the lattice spacing.}. Since we are interested 
only to large times and we know that the systems is trying to 
decrease its local energy $-\frac{m^{2}}{2}\phi^{2}+\frac{g}{4}\phi^{4}$
as well the elastic ones $\frac{1}{2} (\nabla \phi)^{2}$ we expect
(and it can indeed be shown)
that at large times 
\[
<\phi^{2} ({\bf x},t)>=\Delta \int \frac{d{\bf k}}{(2\pi)^{3}}\exp \left(-2 k^{2}t+2 \int_{0}^{t}a (t')dt' \right)
\]
converges to a finite value (actually equal to $m^{2}/g$). But this is
possible only if 
\[
2 \int_{0}^{t}a (t')dt'\simeq 3/2 \log (t/t_{0})\qquad
t_{0}= (\Delta m^{2}/g)^{2/3}/8\pi 
\]
at large times, i.e. $a (t)\simeq \frac{3}{4t}$ for $t>>1$.
Now that we know the expression of $a$ we can compute the space time
correlation function at large times:
\begin{equation}\label{c}
<\phi ({\bf x},t)\phi ({\bf x'},t')>=\frac{m^{2}}{g}\left(\frac{4tt'}{(t+t')^{2}}
\right)^{3/4}\exp \left(-\frac{({\bf  x}-{\bf x'})^{2}}{4 (t+t')} \right)
\end{equation} 
This expression is very interesting and deserves different remarks.
First, for any finite and fixed $({\bf  x}-{\bf x'})$, it becomes at
large times simply a function of $t'/t$. 
Thus, we have found that the correlation between two points, as far as
they could be in space and time, converges to the value
$m^{2}/g$ when $t',t\rightarrow \infty$ {\it provided that their distance in
space and time is kept fixed when the large time limit is taken},
i.e. if one looks to a finite region on a finite time this region 
will be typically on one of the two state $\pm \sqrt{m^{2}/g}$.
On the other hand, for fixed but very large time $t,t'$ the
magnetization density correlation always fall to zero over distance
$|{\bf  x}-{\bf x'}|\propto \sqrt{t+t'}$. This means that, at time t,
the typical size of the regions in the states $\pm \sqrt{m^{2}/g}$ is 
$L (t)\propto \sqrt{t}$. Furthermore, for fixed $|{\bf  x}-{\bf x'}|$,
the magnetization density correlation always falls to zero over a time
separation $t-t'$ {\it which is larger than $t'$} (we assume $t'<t$).
This means that the time it takes to the system to decorrelate from 
its configuration at time $t'$ is of the order of $t'$ itself, i.e. the age
of the system is the characteristic timescale for the dynamical
evolution: the older is the system, the slower is its dynamics.

Thus we recover what discussed in the
introduction and plotted in Fig. \ref{fig2}: the systems is decomposed in
domains of size $L (t)$ (that is the growing correlation I talked
about in the introduction), for which we have found a growth law $L (t)\propto \sqrt{t}$. In the
interior of each domain the system is in one of the two state $\pm
\sqrt{m^{2}/g}$. However, if one waits a time of the order of the age
of the system a point ${\bf x}$ has been swept by different domain
walls and thus the magnetic correlation has been lost.

In this approximate treatment of this simple model we find the
essential ingredient of the coarsening/scaling picture of the aging dynamics.
On long time-scales the system is broken up in domains which grow
and the dynamics is self-similar provided the length is appropriately
rescaled by $L (t)$. This length $L (t)$
is naturally interpreted as the typical size of the domains and its 
rate of growth can be found from general arguments \cite{Bray}. 
In this case $L (t)\propto \sqrt{t}$ but the growth rate may be
different for different systems \cite{Bray}. For example, 
if we consider the Ising model but with a dynamics that conserves the
magnetization, for example the Kawasaki dynamics \cite{Bray}, then 
$L (t)\propto t^{1/3}$. However, because of the self-similarity of the
dynamics, one generically expects that the correlation function can be
written in a scaling form 
\[
<\phi ({\bf x},t)\phi ({\bf x'},t')>=f \left(\frac{|{\bf  x}-{\bf
x'}|}{L (t)},\frac{L (t)}{L (t')} \right)\qquad .
\]
This is indeed the case in our approximate treatment and 
we refer to \cite{Bray,ROE} for a more general discussion.

\begin{figure}
\centerline{\epsfxsize=10cm
\epsffile{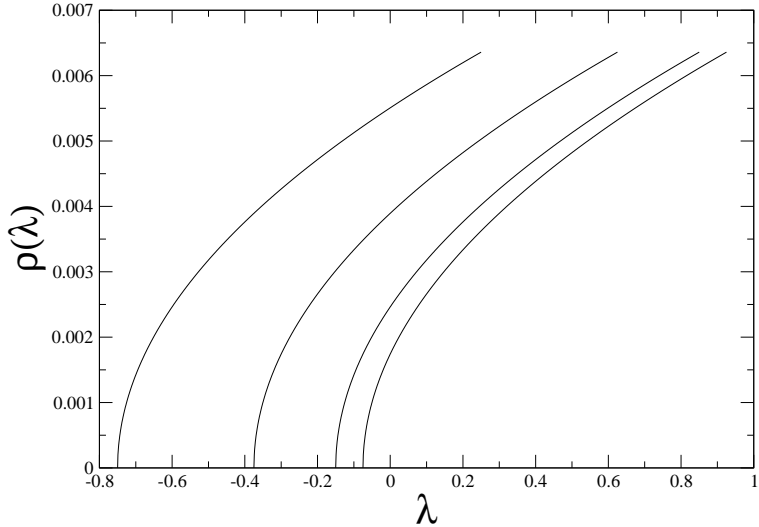}
}
\vspace{.2cm}
\caption{\label{fig3}
$\rho_{t}(\lambda )$ as a function of $\lambda $ for $t=1,2,5,10$
(from left to right). For simplicity $N$ is put equal to one.
}
\end{figure}

Another way to think about aging and glassy dynamics is the one 
coming from the solution of mean-field disordered systems very much
based on the energy landscape (or more precisely free energy but here it does not matter
because $T=0$). 
Although we postpone a more technical discussion to the next section, in the
following we shall introduce the main ideas \cite{ROE,Laloux}.
Let us start with some general remarks. \\
The equation (\ref{t0})
consists in a gradient descent in the energy landscape from a high
temperature configuration toward the ground state configurations (that
are never reached on finite time scales). During this gradient descent
the velocity of the system decreases and tends
to zero in the infinite time limit. This rather natural results 
can be checked noticing that the square of the absolute value  
of the velocity in the energy landscape $\vec{v}=-\nabla F
=-\frac{\delta F}{\delta \phi
({\bf x})}$ equals $-\frac{dF}{dt}$. Since on general grounds one expects that
one time quantities, in particular $F$, tends toward a well defined value
in the infinite time limit (indeed one can check that this is the case
within our Hartree approximation) we obtains that $\frac{dF}{dt}$ and
the velocity has to tend to zero in the infinite time limit.\\
Another important quantity other than the velocity that is useful 
to track the system evolution is the the energy Hessian evaluated at
the configuration reached at time $t$, in particular its density of 
eigenvalues. Since the system is trying to reach the completely stable 
states $\pm \sqrt{m^{2}/g}$ we expect that the fraction of 
negative eigenvalues becomes smaller and
smaller at large times and vanishes at $t=\infty$. 
Let us check this assumption within our approximate treatment in which
the Hessian is diagonal in Fourier space: 
\[
H=\frac{\delta F}{\delta \phi
({\bf k})\delta \phi
({\bf k'})}=\left[{\bf k}^{2}-m^{2}+g<\phi ({\bf x},t)^{2}> \right] \delta({\bf k}+{\bf k'}).
\] 
Computing the density of states is an easy job and one finds:
\begin{equation}\label{rho}
\rho_{t}(\lambda )=\frac{1}{4\pi^2N}\sqrt{\lambda +m^{2}-g<\phi ({\bf
x},t)^{2}>}\quad .
\end{equation}
where $N$ is a normalization factor \footnote{Note that
 since there is an underlying lattice there is also a 
cutoff in the integration over ${\bf k}$. Therefore $N$ is cut-off 
and lattice dependent. Also $\rho(\lambda)$ is in reality cut-off and 
lattice dependent except at its left edge where is given by 
eq. (\ref{rho}).}.
Since $m^{2}-g<\phi ({\bf x},t)^{2}>=a (t)\simeq 3/4t$ we find
that the density of states of the Hessian has at its left edge 
a square root form starting at a negative value 
$\lambda_{min}\simeq - 3/4t$ which
indeed approaches slowly zero at infinite times, see Fig \ref{fig3}.
Finally, the last piece of information that we need is that the velocity 
always points toward directions corresponding to small
eigenvalues of the Hessian. This can be checked noticing that when 
the energy decreases slower than exponentially to its infinite time
value (a quite general behavior in aging system that is true in our case):
\begin{equation}\label{}
\frac{\frac{d^{2}F}{dt^{2}}}{\frac{dF}{dt}}=-2\frac{\nabla F\cdot H\cdot
\nabla F}{(\nabla F)^{2}}\rightarrow 0 
\end{equation}
when $t\rightarrow \infty $. Hence, decomposing this identity along
the eigenvectors of $H$ and calling $v_{\mu }$ the components of
$\nabla F$ on the eigenvector $\mu $ we find:
\[
\frac{\sum_{\mu }v_{\mu }^{2}\lambda_{\mu }}{\sum_{\mu}v_{\mu }^{2}}\rightarrow 0
\]
which shows that indeed the velocity always points toward
directions corresponding to small (vanishing) eigenvalues of the Hessian.\\
Putting all these pieces of information together, we obtain an energy
landscape picture of aging in which the system makes a gradient
descent in the energy landscape and at long times it moves slowly and slowly
(the absolute value of the velocity is going to zero) along channels
with many uphill directions and a low density of 
downhill directions. The system velocity points towards the
directions over which the energy landscape is flatter. 
The longer is the time the smaller is the fraction of downhill
directions and they are more and more flat. Thus it takes
longer to the system to move and go downhill.  
This never-ending descent in the energy landscape dominated 
by an increasing flatness of the 
landscape is the main picture behind aging obtained from the exact
solution of the dynamics of different mean-field models \cite{ROE}.\\
Finally, let me comment on how things change if the final temperature
after the quench is not zero (but still less than $T_{c}$). 
In this case inside the domains the system has a pseudo-equilibrium dynamics 
that takes place on a finite timescale. For example, the local spin-spin
correlation function will be approximatively a function of $t-t'$ for 
$t-t'<<t'$ (and $t'>>1$) that converges to a plateau $m_{i}^{2}$ 
on a finite timescale where $m_{i}$ is the local magnetization. 
The aging regime shows up only on timescales such that $t-t'\propto
t'$. It corresponds to the exit from the plateau and can still be 
described in the same way discussed above.
\section{Disordered systems}\label{disorder}
In the following I will discuss in some detail the theoretical
approaches introduced to explain the aging dynamics.
I will focus mainly on disordered systems for historical reasons and
simplicity and I will comment in the conclusion on the generalization 
to non-disordered glassy systems.
\subsection{Mean-Field Theory of Aging}\label{MFT}
\begin{figure}[ht]
\centerline{
\psfig{file=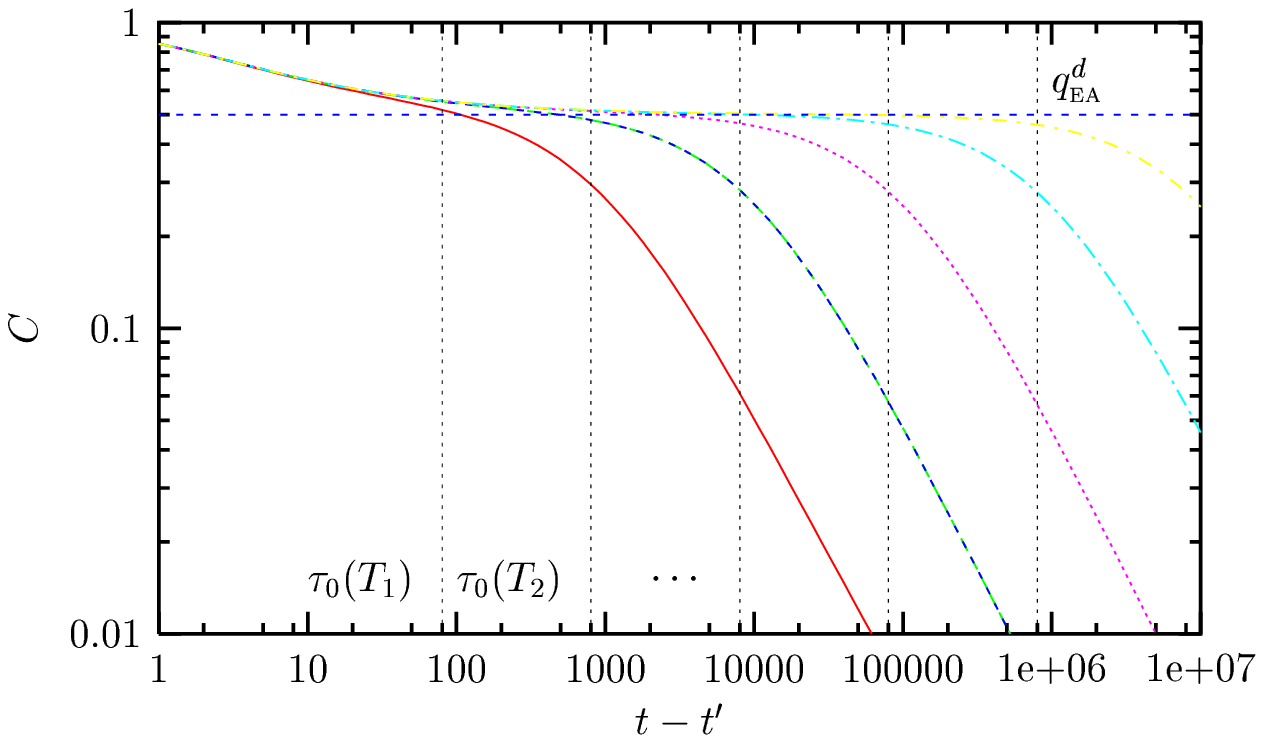,width=8cm}
\psfig{file=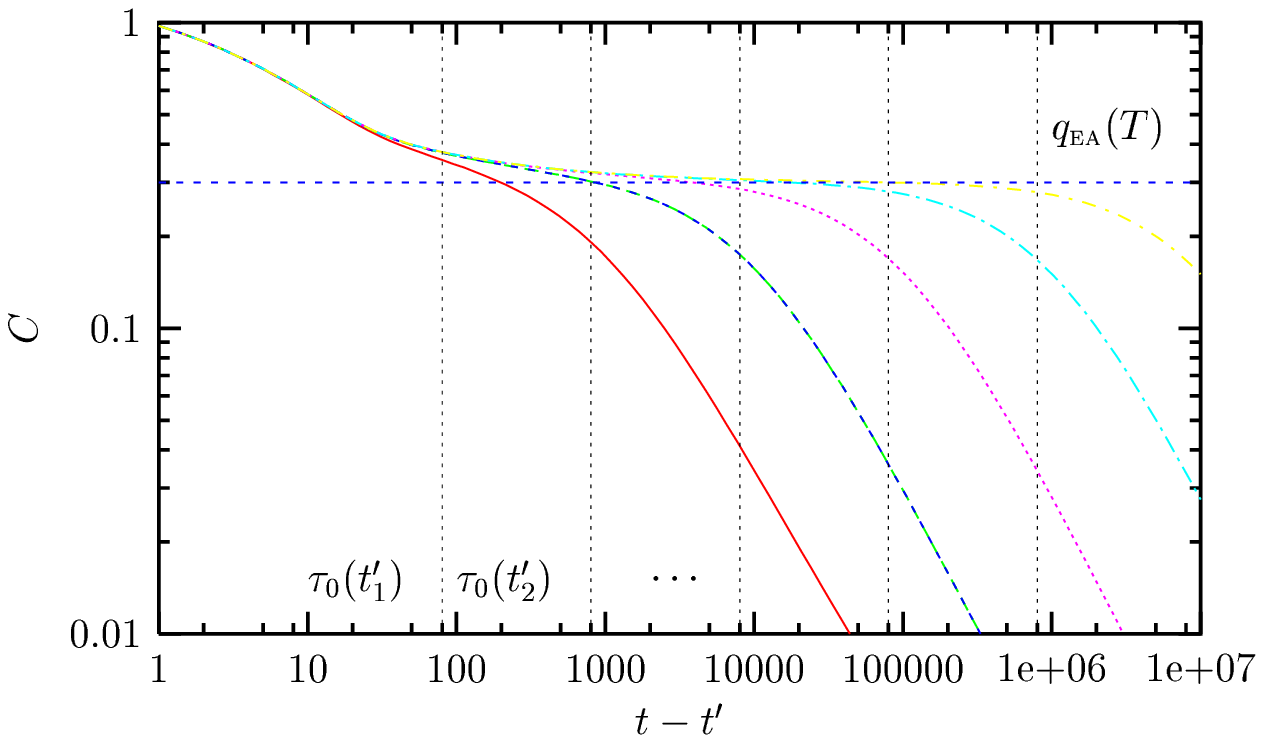,width=8cm}
}
\vspace{-0.5cm}
\caption{Left: Sketch, for a p-spin model, of the decay of the stationary 
correlations in the high $T$
phase close to $T_d$, $T_1 > T_2 > \dots$ (from left to right). Right: Sketch,
for a p-spin model, of the decay of
the aging 
correlations in the low $T$ phase, at fixed $T< T_d$, $t'_1< t'_2<
\dots$ (from left to right).} 
\label{fig4}
\end{figure}

The study of the equilibrium and off-equilibrium dynamics
of mean-field disordered systems revealed that a very rich and 
interesting behavior can be found within the mean-field approximation.
Generically, it has been found that these systems displays a dynamical
transition at a temperature $T_d$ at which the relaxation time diverges.
A quench to a temperature less than $T_d$ leads to an off-equilibrium behavior
and aging that persist forever if the thermodynamic limit has been taken
from the beginning. The way in which the dynamical transition affects
the equilibrium dynamics for $T>T_d$ is discussed in the Lectures
of A. Cavagna \cite{Cavagna} and D. R. Reichman \cite{Dave} and I will not discuss it. 
I would like just to point out
that two quite different behaviors may take place. In mean-field
disordered systems having a spin-glass transition, like for example
the Sherrington-Kirkpatrick model, the dynamical transition temperature
coincides with the temperature at which the equilibrium glass transition
takes place. Instead, for systems like the p-spin disordered 
model, which have an equilibrium transition conjectured to be related
to the structural glass transition, the dynamical transition temperature
is larger than the equilibrium one. Furthermore, contrary to the first
case, the transition has a (partially) first order character since a plateau 
strictly different from zero emerges in the correlation functions  
at $T_d$, see the left plot in Fig \ref{fig4}.\\
Historically, the mean-field theory of aging has been developed analyzing
the off-equilibrium dynamics of completely connected models. 
An example is the p-spins model whose Hamiltonian is
\[
H=\sum_{1\le i_1\le ...\le i_p}J_{i_1,...,i_p}s_{i_1}\cdots s_{i_p}\quad .
\]
and the couplings $J_{i_1,...,i_p}$ are independent Gaussian variables
with zero mean and variance $p!/2N^{p-1}$.
The dynamics of this systems can be fully analyzed 
if the degrees of freedom are variables $s_i$ subjected
a global spherical constraint $\sum_{i=1}^N s_i^2=N$. The usual
strategy is to take a Langevin dynamics for these (spherical) models
and write down the Schwinger-Dyson equations for the correlation $C$
and response functions $R$:
\[
C(t,t')=\frac 1 N \sum_i \overline{<s_i(t)s_i (t')>} \qquad 
R(t,t')=\frac 1 N \sum_i \frac{\overline{<s_i(t)>}}{h_i(t')},
\]
where $h_i(t')$ is a local magnetic time-dependent field coupled to $s_i$
and the overline means the average over disorder. 
A great simplification due to the completely connected character of these
models is that the self-energy can be written as a simple polynomial
function of correlation and response. More specifically, the general
expression one finds is \cite{ROE}:
\begin{eqnarray}\label{sd1}
\int_0^{+\infty} dt_1 R_0^{-1}(t,t_1)C(t_1,t')-2TR(t',t)&=&\int_0^{+\infty}
dt_1 \Sigma_C(t,t_1)R(t',t_1)+\\&&+\int_0^{+\infty}
dt_1 \Sigma_R(t,t_1)C(t_1,t')\nonumber\\
\int_0^{+\infty} dt_1 R_0^{-1}(t,t_1)R(t_1,t')&=&\delta(t-t')+\label{sd2}\\&&+\int_0^{+\infty}
dt_1 \Sigma_R(t,t_1)R(t_1,t')\nonumber
\end{eqnarray}
where $R_0^{-1}$ is the inverse of the bare response function. For example
in the case of the p-spin spherical model $R_0^{-1}(t,t_1)=[\partial_{t}+
\mu (t)]\delta(t-t_1)$ where $\mu(t)$ is a spherical Lagrange multiplier
determined enforcing the spherical constraint $C(t,t)=1$. 
The self-energies are simple functions of the
propagators and read: 
\begin{eqnarray}\label{sigma}
\Sigma_C(t,t')=\sum_n c_n [C(t,t')]^n \qquad 
\Sigma_R(t,t')=\sum_n nc_n [C(t,t')]^{n-1}R(t,t')
\end{eqnarray}
and the coefficient $c_n$ depends on the particular model one is focusing on.
For example in the case of the p-spin spherical model all the $c_n$ are zero
except $c_{p-1}=p/2$, i.e. $c_{n}=\delta_{n,p-1}p/2$.\\
At this point it is easy to understand that it is possible to derive 
these type of equations not only within the exact solution of completely
connected models but also within self-consistent approximate treatment 
of finite dimensional systems. Indeed in the case of Langevin
dynamics one can set up a field theory in which the  
self-energy is written as a sum of diagrams constructed with the vertices
of the theory and using the true propagator as line. The equation
(\ref{sigma}) is in general equivalent to retain only the first self-consistent
diagram and can be often justified in terms as a large $N$ expansion in
$1/N$. This also makes it clear why this 
the type of eqs. introduced above can be found also for 
non-disordered systems. More complicated functional forms have been also
partially studied in the literature but I will not discuss it further.\\
There are two methods for analyzing the eqs. (\ref{sd1},\ref{sd2}): a complete
solution can be obtained by integrating them numerically but much information
can be gained already analytically. In the following I will just make a 
summary of the main findings.\\
For $T>T_d$ correlations and responses converge at large times toward a
time translation invariant regime (TTI), i.e. they depend only on the time difference $t-t'$ and 
furthermore they are related by the fluctuation-dissipation relation
$R(\tau)=-C'(\tau)/T$. Hence, the system is at equilibrium. 
Approaching $T_d$ the correlation function develops a plateau for 
discontinuous spin models and 
takes more and more time to attain the equilibrium TTI regime, see Fig. \ref{fig4}.
Instead for $T<T_d$ the correlation and response never reach the equilibrium 
and a TTI regime. In simple cases, for example in the case of the p-spin
spherical model, there are two time sectors: a pseudo-equilibrium one
that corresponds to taking the limit $t'\rightarrow \infty$ with $t-t'$ fixed. 
In this regime the correlation and response become indeed TTI and are related
by the fluctuation dissipation relation (as for the domain growth
discussed in the previous section). The
aging regime corresponds to $t\rightarrow \infty$ with $t'/t$ fixed 
in which the correlation and response function become functions
of $t'/t$. The correlation function equals the Edwards-Anderson
parameter $q_{EA}$ for $t'/t=1$ and approaches zero in the limit $t'/t=0$
(I take $t'<t$), see Fig \ref{fig4}. 
Note that in more complicated case the $t'/t$ is replaced by $h(t')/h(t)$
with $h$ a model dependent function. It may also happen, in particular 
for continuous models that there is an infinite sequence of time sectors, 
each one of them determined by a ratio $h_i(t')/h_i(t)$. \\
Finally, a striking property of this
off-equilibrium regime is that the correlation and response are still related
by a fluctuation-dissipation like relation (FDR) in which, however,
the bath temperature is replaced by an effective temperature 
$T_{eff}$. In the case of different 
time sectors there is an effective temperature for each one of them.
The interpretation of $T_{eff}$ as an effective temperature has been
justified in different papers. Furthermore, it has also 
been shown that $T_{eff}$ can
be related in some cases to the Parisi function $x(q)$ introduced in the
study of the thermodynamics. This topic has generated an enormous
amount of interest: many numerical simulations and some experiments
have been performed in order to check the existence of this
off-equilibrium relation between correlation and response. A very
popular way to plot data to check for a generalization of FDR
consists in plotting the integrated response $\chi
 (t,t_{w})=\int_{t_{w}}^{t}R (t,t')dt'$ as a
function of the correlation $C (t,t_{w})$. If the fluctuation
dissipation relation $R=-C'/T_{eff}$ is verified (in the limit $t_w
\rightarrow \infty$) one should find 
a straight line with slope $-1/T_{eff}$. 
\begin{figure}[ht]
\centerline{
\psfig{file=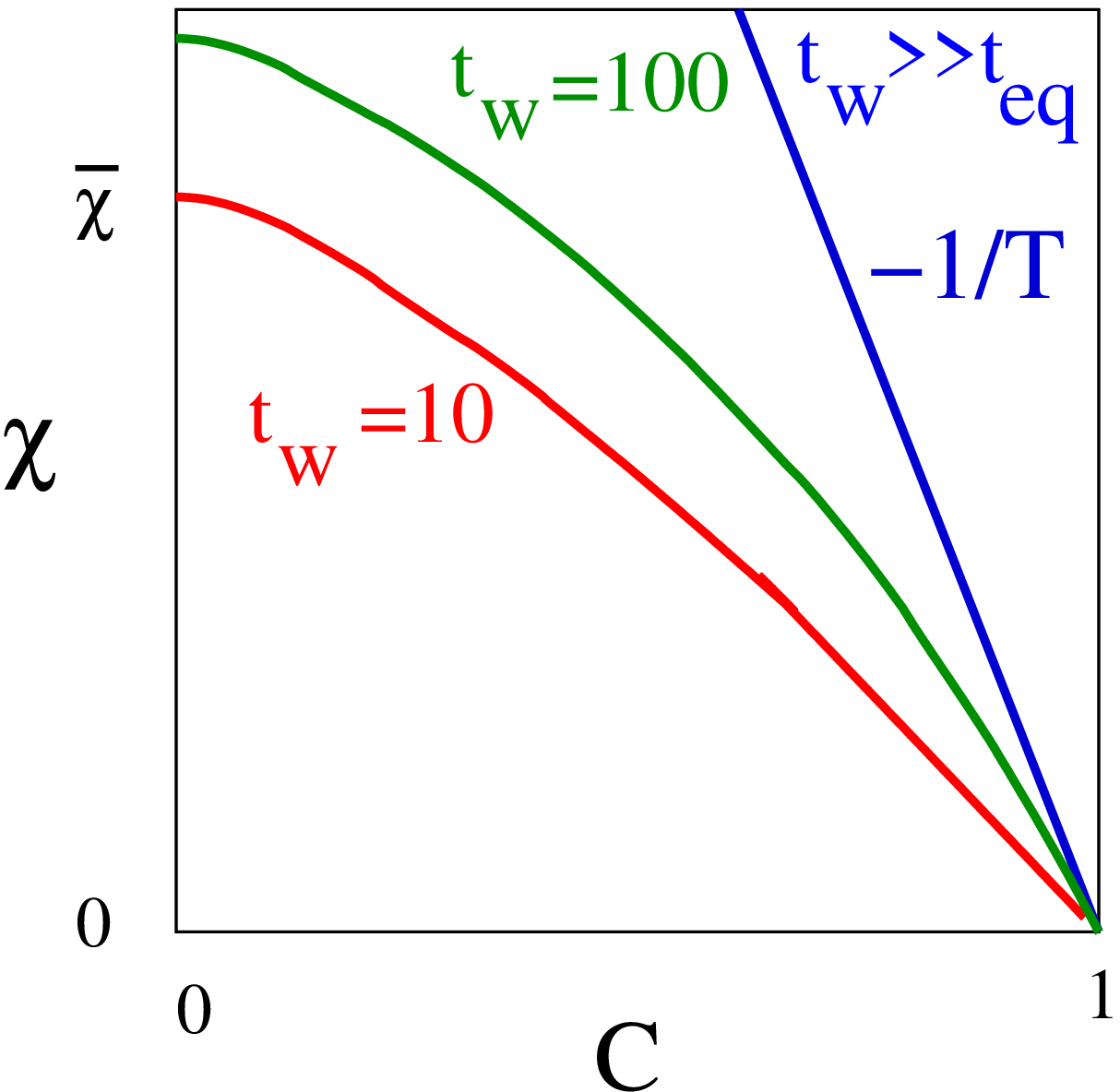,width=5cm}
\hspace{1cm}
\psfig{file=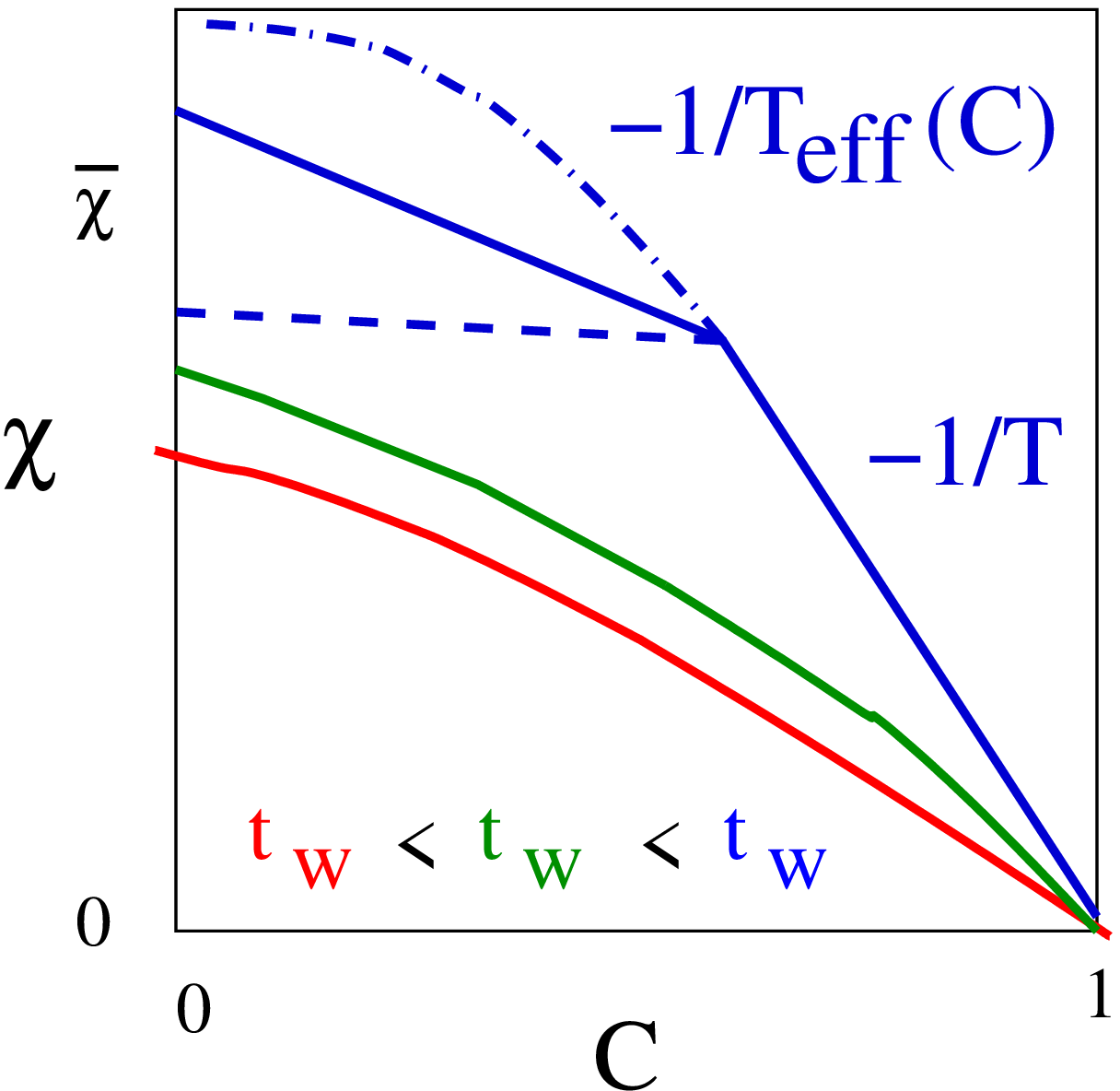,width=5cm}
}
\caption{\label{fig5}
The asymptotic behavior of the integrated linear response against the 
correlation in a parametric plot, for fixed waiting time and using $t$ 
as a parameter. Left: behavior for a system relaxing to equilibrium.
Right: behavior in a slowly relaxing system out of equilibrium.
The dot-dashed line corresponds to a case with an infinite number of
timescales $h (t)$ and $T_{eff}$. The continuous line with a breaking
point corresponds to a systems, like the p-spin spherical model, with
only one $T_{eff}$. The dotted line corresponds to the domain growth
in the Ising model. Finally the other curves correspond to waiting
times far from the asymptotic regime.}

\end{figure}
In Fig. \ref{fig5} we plot on the left a typical FDR plot for a system 
converging to equilibrium. In this case the curves, for different
$t_{w}$ evolves toward a straight line. Instead for mean-field glassy
systems off-equilibrium a rather different behavior has been found,
see the right part of Fig. \ref{fig5}. The dot-dashed line 
corresponds to a case with an infinite number of
timescales and $T_{eff}$. The continuous line with a breaking
point corresponds to a systems, like the p-spin spherical model, with
only one $T_{eff}$. The dotted line corresponds to the domain growth
in the Ising model $T_{eff}=\infty $. 
The other curves correspond to waiting times far from 
the asymptotic regime.
A detailed discussion about $T_{eff}$ can be found in \cite{ROE}. 
I will just make some other comments in the conclusion.\\
Finally, let me discuss the physical interpretation of this aging dynamics.
As anticipated in the previous section the interpretation that comes
from many different analytical and numerical studies is based on a
never-ending descent in the (free-)energy landscape\footnote{In reality only at
zero temperature this interpretation is fully justified because
otherwise the dynamics cannot be mapped exactly into an evolution
within the free-energy landscape.}. Indeed the aging after quench at zero temperature
is very similar to his finite temperature counterpart. Instead a very different behavior
would arise if aging was due to activated hopping over (free-) energy 
barriers. The interpretation is exactly the same outlined in the previous
section on coarsening except that now the eigenvectors of the Hessian change during the
time evolution. Another important new element with respect to domain
growth in the Ising model is that for these mean-field glassy systems
there are an exponential number (in the systems size) of thermodynamic
states. It is generically believed and indeed it has been shown
in simple cases that the system ends up, at long times, aging
very close to the states 
that are marginally stable, i.e. that have a vanishing fraction of
flat directions (at least after a very slow quench). Finally,
in the case of discontinuous spin glasses, the effective temperature 
has been related to the complexity $S (f)$, 
i.e. the log degeneracy of states with a given free-energy $f$,
as $1/T_{eff}=\partial S/\partial f$ where the derivative is evaluated
at the free energy density of the marginally stable states.\\
I will not continue further the discussion of the mean field theory
of aging. I refer to more detailed review for deeper analysis
\cite{ROE} and to the conclusion for a general discussion.
\subsection{Activated dynamic and scaling}\label{activated}
Another way to tackle and describe the aging dynamics that has been developed
in the last twenty years is based on a growing correlation length,
scaling and renormalization group ideas. The physical picture that one
has in mind is similar to the one discussed for the aging dynamics of the
Ising model. After a quench to the low temperature phase domains
separating the two (or more) phases grow. However, because of quench disorder,
the domain walls have to overcome (free-)energy barriers in order to
grow. The scaling hypothesis is that at
a timescale $t$ the typical linear domain size is
$\xi (t)$ and that correlation functions and responses at different
times can be written in terms of a unique scaling function up to a
length rescaling. For example, as for the non disordered Ising model
the spin-spin correlation function is expected to read:
\begin{equation}\label{scaling}
<\phi ({\bf x},t)\phi ({\bf x'},t')>=f \left(\frac{|{\bf  x}-{\bf
x'}|}{\xi (t)},\frac{\xi (t)}{\xi (t')} \right)\qquad .
\end{equation}
On this lengthscale, $\xi(t)$, the typical (free-)energy barrier that domains
have to overcome in order to grow is assumed to scale like $\xi^{\psi}$. As a consequence,
using the Arrhenius law relating time and energy barrier one finds
that $\xi (t)\propto (T\log t)^{1/\psi}$, i.e. a very slow activated
aging dynamics. Note that the previous expression is correct only in
the regime $\frac{\xi (t)}{\xi(t')}\neq 1$. 
Also in this case one expects a pseudo-equilibrium regime,
corresponding formally to $t,t'>>1$ with $\frac{\xi (t)}{\xi(t')}=1$
and physically to the equilibrium relaxation in the regions inside the domains.
A systems in which this description is believed to be correct is the
Random Field Ising model, RFIM, (ferromagnetic Ising model with independent
random fields on each site). This system has a ferromagnetic phase
transition for a not too high value of the variance of the random fields (their mean is
zero). After a quench at low temperature domains separating
the positive magnetization phase from the negative magnetization phase 
grows following the laws discussed above.
This description of the aging dynamics has been proposed for many
different disordered systems like disordered ferromagnets, spin
glasses, etc\dots . In some cases, like for spin-glasses, this description
of the aging dynamics is still matter of debate. Notice also that, at least
superficially, it is very different from the mean-field one described in the
previous section which is not based at all on activated dynamics. \\
In the following I shall present a sketchy analysis of the
off-equilibrium dynamics of the 1d RFIM \cite{Cecile}. This model is an
instructive example because the above heuristic predictions can be
indeed obtained analytically and their physical content can be easily
grasped.
The Hamiltonian is:
\[
H=-J\sum_{i=1}^{N-1}s_{i}s_{i+1}-\sum_{i=1}^{N}h_{i}s_{i}
\]
where $h_{i}$ are independent, say Gaussian, random variables with
zero mean and variance $g=\overline{h^{2}_{i}}$. At zero temperature,
the Imry-Ma argument implies that the region of $+$ and $-$ have
a typical size $L_{IM}=4J^{2}/g$. The reason is simple: over a length
$l$ the energy gain obtained aligning the spins with the random fields
is $\propto -\sqrt{gl}$ whereas the energy loss for creating a domain
wall between a region of $+$ and $-$ spins is $2J$. These two
contributions match precisely at $l=L_{IM}$. Thus, we find
that the RFIM in one dimension has of course no phase transition;
however as long as the typical distance between domains during aging
is less than $L_{IM}$ (or its finite-temperature counterpart) 
the system does not know it and the aging behavior
is expected to be described by the activated dynamics scaling scenario
described before. Note that this is also an example of interrupted aging
in which if one waits long enough the system eventually equilibrates.\\
The way to tackle the aging dynamics of the 1D RFIM is to rewrite its
Hamiltonian in terms of positions of domain walls. Let me call $a$ and
$b$ respectively the domains $+|-$ and $-|+$, $N_{a}$ and $N_{b}$
their number, and $a_{i}$ and $b_{i}$ the position of the ith domain
of type $a$ and $b$. Using this variables one can write the Hamiltonian
as
\[
H=-J (N-1) -\sum_{i=1}^{N}h_{i}+2J (N_{a}+N_{b})+\sum_{i}^{N_{a}}V (a_{i})-\sum_{i}^{N_{b}}V (b_{i})
\]
where $V (x)=-2\sum_{i=1}^{x}h_{i}$. Furthermore, a Glauber
dynamics on the spins (i.e. with a transition rate $w (s_{i}\rightarrow
-s_{i})=e^{-\beta \Delta E} /(e^{-\beta \Delta E}+e^{+\beta \Delta E})$)
implies a reaction-diffusion dynamics for the domains in which a couple domains can be created
on two neighboring sites with energy cost $4J\pm h_i$, diffusion takes place with an energy
cost $\pm 2h_{i}$ and the domains annihilate on neighboring sites with
an energy cost $-4J\pm h_{i}$. We are interested in the aging dynamics after
a quench to a very low temperature $T<<J$ and with $g<<J^{2}$. This is
the case in which $L_{IM}>>1$ and the activated dynamics scaling
picture should hold. In this case the effective dynamics for the
domains is: $a$ domains diffuse within the external potential $V$,
$b$ domains diffuse within the external potential $-V$ and $a$ and $b$
annihilates when they meet.
The activated dynamics scaling picture now becomes very concrete. At 
time $t$ domains are at a typical distance $\xi(t)$. This length is
just the typical distance $d_{S} (t)$ on which a particle diffuses in the interval 
of time $t$ in the random potential $V$ (or $-V$, the sign does not
matter). The main reason is that on
lengthscales smaller than $d_{S} (t)$ there should be just one domain
because if there were others then they would have meet and therefore
they have been annihilated on timescale $t$. Instead on lengthscales
larger than $d_{S} (t)$ all the domains that were present before the quench
are still there, thus $\xi(t)=d_{S} (t)$. \\
The problem of determining $d_{S} (t)$  and analyzing the motion
before annihilation of $a$ and $b$ 'particles' is very well-known 
because is a discrete version of the Sinai Model: a particle diffusing with a Langevin
dynamics in a Gaussian independent random force field. Indeed
integrating the force to get the potential one finds precisely 
$V (x)=\sum_{i=1}^{x}f_{i}$ where $f_{i}$ are independent Gaussian random
variables. It has been proved by Sinai that $d_{S} (t)\propto 
(T\log t)^{2}$. There is also a simple argument to heuristically get  
that $\psi=2$: the potential $V (x)$ in an interval of length $d$ is
like a Brownian motion (in which $V$ is the position and $x$ the time),
thus the typical excursion of $V$ around its boundary values 
is $\sqrt{d}$. Hence, using the Arrhenius law we get
precisely $d_{S} (t)\propto (T\log t)^{2}$.\\
Finally, the other nice thing about the dynamics of the 1D RFIM is that it can be
really fully analyzed within a real space renormalization group procedure 
that provides an analytical derivation of the activated dynamics scaling
picture and of many other results \cite{Cecile}. This RG procedure has been
introduced by Dasgupta and Ma and for disordered quantum spin
chains and it has been developed and put on a firmer theoretical ground
by D.S. Fisher and subsequently applied to many different systems
\cite{Cecilereview}.\\
I will not continue further the presentation of this very appealing 
scenario for the aging dynamics and I refer to reviews for other
details \cite{Cecilereview,ROE} and to the conclusion for a further discussion.
\subsection{Trap Model}\label{Trap}
\begin{figure}[hbt]
\[
\includegraphics[width=10cm]{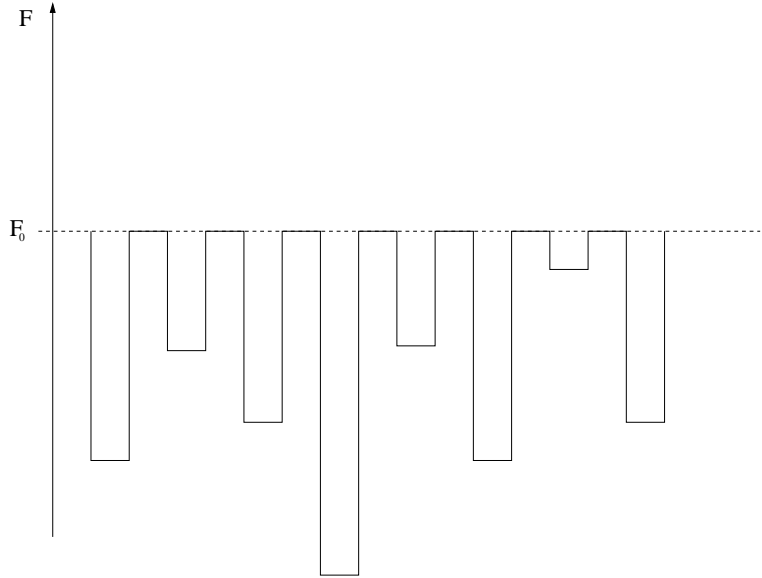}
\]
\caption{Schematic representation of the configuration space for the
trap model.}
\label{fig6}
\end{figure}

The trap model for aging dynamics has been introduced by Bouchaud in
\cite{Trap} and extensively developed since then \cite{ROE}.
The trap model is not a microscopic model. It is however very
interesting because on the one hand it has been used 
and it is still used a lot on a phenomenological level to interpret 
results of experiments and simulation and, on the other hand, it provides a coarse
grained description of different microscopic models. \\
The simplest version \cite{Trap} is defined as follows: there are $N$
possible states plotted as wells in Fig. \ref{fig6}. The dynamical
evolution is determined by stochastic jumps: each well acts as a trap
for the system. The trapping time is determined by the Arrhenius law:
$\tau_{\alpha } =\tau_{0}\exp (+(F_{0}-F_{\alpha })/T)$ where
$F_{\alpha }$ is the free energy of the ``state'' (or well, trap)
$\alpha $ and $F_{0}$ is a threshold free-energy state, see
Fig. \ref{fig6}.
Once the system is escaped from a trap, it falls
completely at random into another trap. The rationale behind
these rules is that generically for disordered systems there exists 
many different metastable states which, at least in some regimes, 
are surrounded by large energy barriers. The complete absence of
geometry is of course the simpler assumption one can do. More 
complicated structures of the configuration space has been considered.
The other essential ingredient of the model is the distribution of 
free-energies that is taken exponential with a certain parameter $x$:
\[
P (F)=x/T\exp (x (F-F_{0})/T).
\]
This choice is motivated by the presence of a 
similar distribution in mean-field
disordered systems like the Random Energy Model (in which $x=T/T_{g}$)\cite{ROE} and, on
general grounds, by the extreme value statistics for deep free-energy
states \cite{ROE}. \\
The resulting distribution of trapping times $\psi (\tau )$ reads:
\[
\psi (\tau )=x \frac{\tau_{0}^{x}}{\tau^{1+x}}\quad .
\]
The important feature of this law is that the average trapping time
$<\tau >$ is infinite for $x<1$, i.e. $T<T_{g}$ if we take $x$ as
proportional to the temperature. In this regime the system, taken at
time zero uniformally distributed over the traps, never reaches the
equilibrium and ages forever. In this case the nature of the aging
dynamics is particularly clear \cite{Trap}: the time $t_{w}$ after $n$ jumps
equals $\sum_{i=1}^{n}\tau_{i}$ where $\tau_{i}$ are independent
variables with probability law $\psi (\tau )$. It is a well-known
result that since the first moment is infinite the central limit
theorem does not apply, $t_{w}$ is distributed with a Levy Law
and its typical value is of the order of the maximum trapping time
encountered during the evolution. At a given waiting
time $t_{w}$ the sum $\tau_{1}+\tau_{2}+\dots $ is dominated by its largest
term which is of the same order of $t_{w}$. In this case it is very
clear that the system ages because the relaxation time is set by the
age of the system itself.\\
I will not present further results on the trap model and I refer to
\cite{Trap,ROE} for a wider presentation. I would like just to stress that 
the Trap Model provides a coarse grained description of the Sinai
model in a biased potential \cite{Cecilereview} as well as the 
aging dynamics (on extremely long times, diverging with the system
sizes) of the Random
Energy Model \cite{Bovier}. These results suggest that it is not only a
phenomenological model and it might play in the future an
important role in the description of the finite dimensional extension
of the mean-field aging.

\section{Outstanding issues}
In this section I would like to point out what are in my mind
some outstanding theoretical issues in aging of glassy systems. 
It will be certainly
amazing to look at this list in a couple of years to see how things
turned out to be very different.\\
{\it \bf  Beyond the mean-field theory of Aging.}\\
{\it Length-scales, Landau Theory and fluctuations.}
As for critical phenomena one can identify three steps 
that have to be followed to analyze the physics of finite
dimensional systems. First, one can solve an infinite dimensional
system. This has been done for aging and these studies indeed
lead to a mean-field theory of aging \cite{ROE}. A second important
step is set up a Landau theory. Although the results still 
are mean-field like (i.e. no fluctuations are taken into account)
this provides hints of why results might be universal (at least in
higher enough dimension) of which type of correlation lengths are
diverging (as we discussed in the introduction if the aging is not
interrupted one expect a diverging length) and it helps to clarify
what are the important fluctuations that one has to take into
account. I think that it is a very important point that will be 
studied in detail in the future, some results have been already
obtained in \cite{ChamonCugliandolo}.\\
{\it Beyond mean-field theory: activated processes}.
For mean-field systems conjectured to be related to structural glasses
the thermodynamic as well as the equilibrium and off-equilibrium dynamics
are expected to change drastically in finite dimension. The main
reasons is that the physics of these models is dominated by the
existence of an infinite number of thermodynamic states and, instead,
it is clear that in finite dimension this exponential proliferation
of thermodynamic states cannot take place. How this is going to change
the physics (statics and, specially, dynamics)? How this translate in
terms of real space structure? What remains of the off-equilibrium
dynamics described in terms of energy landscape picture?
These are certainly very important and pressing questions that, hopefully,
will receive a lot of attention in the near future.\\
{\it \bf  Renormalization group for the activated dynamic scaling.}\\
Although in one dimension there are examples of activated dynamic
scaling that can be derived and fully analyzed analytically within a 
RG procedure the extension to higher dimensions remains an important 
open problem and certainly a major challenge for the
future. Recent interesting results have been already obtained in
\cite{Balents,Scher}.\\
{\it \bf  Generalization of the Fluctuation-Dissipation relation in
finite dimension}\\
This question has already attracted a lot of attention.
In particular one would like to know if and how the FDR is generalized in
finite dimension in comparison with mean-field predictions
and if the resulting $T_{eff}$ can be really
interpreted as an effective temperature. 
Furthermore also the generalization of FDR to {\it driven}
off-equilibrium systems
as for example granular media is an interesting related question. 
Despite many simulations and some experiments have been done these
questions are not completely settled and certainly they will continue
to give rise to many works in the future. \\
{\bf Aging in glassy systems without quenched disorder}\\
There are systems like structural glasses that have a very interesting
slow dynamics that is still a puzzle despite many years of theoretical
and experimental investigations. The understanding of their 
off-equilibrium dynamics is therefore a challenge for the future.\\
From this point of view it is very likely that in the future 
much more attention will be devoted to the off-equilibrium dynamics of simple 
statistical mechanics model of glassy systems as Kinetically
Constrained Models, Lattice Glasses, \dots (see \cite{RitortSollich}
for a review).\\
{\bf Quantum off-equilibrium dynamics}
What about quantum glassiness? This is certainly a very interesting subject
that will probably be the target of many investigations in the future.
To me very interesting questions are: what are the properties of 
off-equilibrium dynamic (aging) close to a quantum critical point?
What are the properties of the aging behavior induced by a quench
(that goes through a quantum phase transition) due to a sudden change 
in the control parameter governing quantum fluctuations?
What is a quantum glass without disorder?\\
Some of these questions have been recently addressed in \cite{Schmalian,Chamon}.\\
I will stop here this list that could go on much further. 
I think I have described already enough challenging problems.
 
\section{Suggested readings}\label{suggested}
Suggested readings that are useful to have a deeper understanding 
of the aging dynamics and the theoretical ideas developed to
understand it are:\\
J. Kurchan and L. Laloux, J. Phys. A {\bf 29}, 1929 (1996)
for the energy descent picture of aging.\\
L. F. Cugliandolo and J. Kurchan, Phys. Rev. Lett. {\bf 71}, 173
(1993) for the aging dynamics in mean field glassy systems.\\
D. S. Fisher and D. A. Huse, Phys. Rev. B {\bf 38}, 373 (1988),
Phys. Rev. B {\bf 35}, 6841 (1987) for the activated dynamic scaling.\\
J.-P. Bouchaud, J. Phys. I France {\bf 2}, 1705 (1992) for the Trap model.\\
This list is clearly not exhaustive.

\vskip 1cm
{\bf  Acknowledgments}

I thank the organizers of the school and conference ``Unifying
Concepts in Glass Physics III'' for the invitation to give these
lectures and C. Monthus for a careful reading of this manuscript.
Along the years I learned a lot about aging dynamics 
from many colleagues and in particular L.F. Cugliandolo and J. Kurchan
that I am very glad to thank.
Finally I acknowledge partial support from the European Community's
Human Potential Programme contracts HPRN-CT-2002-00307 (DYGLAGEMEM).


\begin{thebibliography}{99}

\bibitem{ROE} J.-P. Bouchaud, L. Cugliandolo, J. Kurchan, M.
 M{\'e}zard, in {\it Spin-glasses and
  Random Fields}, edited by A.~P. Young (World Scientific, Singapore,
  1998).\\
  L. Cugliandolo, in {\it Slow relaxations and non-equilibrium dynamics
  in condensed matter}, 
  Les Houches, Session LXXVII, J. L. Barrat, M. Feigelman, J. Kurchan,
  J. Dalibard Edts, Springer-EDP Sciences (2003).\\
 J.-P. Bouchaud in {\it Soft and Fragile Matter}, M.E. Cates
and M. Evans Edts, IOP (2000).

\bibitem{ReviewSaclay} 
E. Vincent, M. Hammann, M. Ocio, J.-P. Bouchaud, L. F. Cugliandolo
cond-mat/9607224, Proceedings of the Sitges conference 
(E. Rubi ed, Springer-Verlag, 1997). 

\bibitem{Bray} A. J. Bray,  Adv. in Phys. {\bf 43} (1994) 357.


\bibitem{Laloux} J. Kurchan and L. Laloux, J. Phys. A {\bf 29}, 1929 (1996).

\bibitem{Cavagna} A. Cavagna, Lecture notes in the same volume.

\bibitem{Dave} D.R. Reichman, Lecture notes in the same volume.

\bibitem{Cecile}  D. S. Fisher, P. Le Doussal, C. Monthus,
Phys. Rev. E {\bf  64}, 066107 (2001).

\bibitem{Cecilereview} For a review see {\it Strong disorder RG approach of random systems} 
C. Monthus and F. Igloi, cond-mat/0502448, Phys. Rep. in press.

\bibitem{Trap} J.-P. Bouchaud, J. Phys. I France {\bf 2}, 1705 (1992).

\bibitem{Bovier} G. B. Arous, A. Bovier, and V. Gayrard
 Phys. Rev. Lett. {\bf  88}, 087201 (2002).

\bibitem{ChamonCugliandolo}  C. Chamon, M. P. Kennett, H. Castillo, L. F. Cugliandolo,
Phys. Rev. Lett. {\bf  89} 217201 (2002); H. E. Castillo, C. Chamon,
L. F. Cugliandolo, M. P. Kennett, Phys. Rev. Lett. {\bf  88}, 237201 (2002).
  
\bibitem{Balents} L. Balents and P. Le Doussal, EuroPhys. Lett. {\bf
65} 685 (2004) and condmat/0408048 to appear in Adv. in Physics.

\bibitem{Scher} G. Schehr, P. Le Doussal, {\it Functional
Renormalization for pinned elastic systems away from their steady
states}, condmat/0501199; Phys. Rev. Lett. {\bf  93}, 217201 (2004).

\bibitem{RitortSollich}  F Ritort and P Sollich, Adv. in Phys. 
{\bf  52} 219 (2003). 

\bibitem{Chamon} C. Chamon, Phys. Rev. Lett. {\bf 94}, 040402 (2005).

\bibitem{Schmalian} H. Westfahl  Jr., J. Schmalian, P. G Wolynes, 
Phys. Rev. B {\bf  68}, 134203 (2003).



\end{thebibliography}
\end{document}